\documentclass[12pt]{article}
\usepackage{amsfonts}
\def\sqr#1#2{{\vcenter{\hrule height.#2pt\hbox{\vrule width.#2pt
height#1pt \kern#1pt \vrule width.#2pt}\hrule height.#2pt}}}
\def\square{\mathchoice\sqr64\sqr64\sqr{4.2}3\sqr{3.0}3}

\def\br{\begin{eqnarray}}
\def\er{\end{eqnarray}}
\def\brn{\begin{eqnarray*}}
\def\ern{\end{eqnarray*}}
\def\er{\end{eqnarray}}
\def\be{\begin{equation}}
\def\ee{\end{equation}}
\def\vt{\vartheta}

\def\la{\lambda}
\title{Gravity on a parallelizable manifold\\
Exact solutions\\
}
\author{
\thanks {\quad   email itin@sunset.ma.huji.ac.il}
Yakov Itin \\
\small {Institute of Mathematics}\\
\small {Hebrew University of Jerusalem}\\
\small {Givat Ram, Jerusalem 91904, Israel}\\
}
\begin{document}  
\setlength{\evensidemargin}{0in}
\setlength{\oddsidemargin}{0in}
\setlength{\textwidth}{6.25in}
\setlength{\textheight}{8.5in}
\setlength{\topmargin}{0in}
\setlength{\headheight}{0in}
\setlength{\headsep}{0in}
\setlength{\itemsep}{-\parsep}
\setlength{\parskip}{\medskipamount}
\renewcommand{\topfraction}{.9}
\renewcommand{\textfraction}{.1}
\newcommand{\ol}{\setlength{\itemsep}{0pt.}\begin{enumerate}}
\newcommand{\eol}{\end{enumerate}\setlength{\itemsep}{-\parsep}}
\newtheorem{THEOREM}{Theorem}[section]  
\newenvironment{theorem}{\begin{THEOREM} \hspace{-.85em} {\bf :} 
}%
                        {\end{THEOREM}}

\newtheorem{DEFINITION}[THEOREM]{Definition}
\newenvironment{definition}{\begin{DEFINITION} \hspace{-.85em} {\bf 
:} \rm}%
                            {\end{DEFINITION}}

\newcommand{\thm}{\begin{theorem}}

\newcommand{\dfn}{\begin{definition}}

\newcommand{\prf}{\noindent{\bf Proof:} }

\newcommand{\ethm}{\end{theorem}}

\newcommand{\edfn}{\bbox\end{definition}}

\newcommand{\eprf}{\bbox}
\newcommand{\beqn}{\begin{equation}}
\newcommand{\eeqn}{\end{equation}}
\newcommand{\wbox}{\mbox{$\sqcap$\llap{$\sqcup$}}}
\newcommand{\bbox}{\vrule height7pt width4pt depth1pt}
\newcommand{\qed}{\bbox}

%
\def\hook{\hbox{\vrule height0pt width4pt depth0.5pt
\vrule height7pt width0.5pt depth0.5pt \vrule height0pt width2pt
depth0pt} }

\newcommand{\bi}[1]{\bibitem{#1}}
\date{\today}

\maketitle
\begin{abstract}
The wave type field equation $\square \vt^a=\la \vt^a$,
 where $\vt^a$ is a coframe field on a space-time,
 was recently proposed to describe the gravity field. 
This equation has a unique static, spherical-symmetric,
 asymptotically-flat solution,
 which leads to the viable Yilmaz-Rosen metric.    
We show that the wave type field equation is satisfied
 by the pseudo-conformal frame if the conformal factor is
determined by a scalar 3D-harmonic function.
 This function can be related to the Newtonian potential 
of classical gravity.
 So we obtain a direct relation between the non-relativistic
 gravity and the relativistic model: every classical exact solution
 leads  to a solution of the field equation. 
With this result we obtain a wide class of exact, static metrics. 
We show that the theory of Yilmaz relates to the pseudo-conformal 
sector of our construction. We derive also a unique cosmological 
(time dependent) solution of the described type. 
\end{abstract}
\section{Introduction}
The study of possible geometrical models of physical reality began 
soon after Einstein had proposed his theory of gravity field - 
general relativity (GR). All these classical field-theoretical
 generalizations of Einstein's theory include some alternation of the
 primordial geometrical structure - the pseudo-Riemannian manifold \cite {G-H},\cite {T-W}.\\
In order to describe the spinorial properties of the matter 
on the curved space-time one needs the existence of an orthonormal 
frame in every point of the manifold \cite {G-H}. This result can be 
formulated mathematically in the form of the Geroch theorem \cite{Ge}:
\thm
A necessary and sufficient condition for a space-time $M$ 
(4D differential manifold with Lorentzian signature) to admit 
a spinor structure is that the orthonormal frame bundle $FM$ 
has a global cross-section.
\ethm
So one needs to consider two different geometrical structures 
on the differential manifold: 
\begin{itemize}
\item \it {frame structure} \rm  (a global cross-section 
                                    of the frame bundle) 
\item \it {metric structure} \rm (a global cross-section 
                       of the second rank tensorial bundle).
\end{itemize}
 These two structures are not complete independent - 
 from one side one needs some metric to define the orthonormality 
of the frame, on the other side the metric can be obtained 
by the coordinate components of the orthonormal frame.\\
A question in the order is:\\
 \it Which of these two structures should be taken 
to play a role of a primary dynamical variable for the gravity field? \rm \\
In the classical Einstein's theory of GR the metric structure
completely describes the gravity field. 
In most of the classical alternative 
theories (from Einstein-Cartan to MAG) \cite {G-H}, \cite {hehl95} 
the frame structure is used to describe the spinorial 
properties of the matter, 
but the pure gravitational sector is described by the metric structure
only.
Therefore, in the modern approach to the gravity-material
system one needs to use two different geometrical structures 
described above. \\
 In \cite{K-I} we make an attempt to eliminate the metric from 
the status of an independent dynamical variable and use the 
frame structure as the only gravity variable. 
So we hope to construct some ``frame gravity'' instead of the 
traditional ``metric gravity''.
 This model  can be considered as one of the teleparallelism
 class gravitational gauge field theories \cite {Hehl}.\\
We begin with the differential manifold $M$ endowed with a
 fixed cross-section of the coframe bundle $FM$ - coframe field
 $\{\vt^a, \ a=0,...,3\}$.\footnote{We use the coframe bundle instead
 of the frame bundle in order to use the exterior algebra technique. 
 It is obvious that all topological condition, such as the Geroch 
 theorem hold true also in this situation.} 
The manifold $M$ is also endowed with the Lorentzian signature 
 of type $(+,-,-,-)$. It means that the Lorentzian scalar product 
 is defined on the (co)tangent vector space in every point $x\in M$.
 However, the metric structure is not defined yet since the
 scalar products in distinct points are not connected.
 In other words we deal with the whole class of
 the Lorentzian signature metrics.\\
The next step is \it{to call} \rm the coframe $\{\vt^a\}$ -
 \it { `` a pseudo orthonormal coframe''}\rm.
 With this ``magic name'' we are able now to fix the metric
 structure of the manifold $M$.
In this way  we also obtain the unique Hodge dual map 
on the algebra of differential forms 
$\Lambda=\sum^n_{p=0}\Lambda^p$. 
We begin with the notation of Hodge dual map
 on an arbitrary vector space.
\dfn
 Let $V^*$ will be a $(n+1)$-dimensional vector space
 with a basis $\{\vt^0,\ldots,\vt^n\}$. 
\newline
\it {Hodge dual map} \rm is an $\mathbb R$-linear map which
 acts on monomial expressions of $ \vt^{a_j}$ in the following way 
$$*(\vt^{a_0} \wedge \ldots \wedge \vt^{a_p}) = 
 (-1)^s\vt^{a_{p+1}} \wedge \ldots \wedge \vt^{a_{n}},$$
where all indices are different and are taken in such an order,
 that the sequence $\{a_0,\ldots{a_p},{a_{p+1}},\ldots,a_{n}\}$ 
 is an even permutation of the sequence $\{0,\ldots,n\}$.
The integer $s$ is:
\be
s=\left \{ \begin{array}{ll}
         0,& \textrm {if 0 is among ${a_0},\ldots,{a_p} $} \\
         1,& \textrm {if 0 is among $a_{p+1},\ldots,{a_n}. $} \\
\end{array} \right.
\ee \edfn
The integer $s$ in the definition above describes the signature of
 the vector space.\\
Now by the linearity the definition can be extended to the algebra
 of the exterior forms on $V$.
The vector space $V$ can be identified with the tangent space of the 
differential manifold $M$ and by the smooth structure on $M$ the 
Hodge dual map can be extended (as a smooth operation) to the whole 
algebraic bundle on $M$.\\
The unique defined metric tensor $g$, which makes the frame $\vt^a$
 to be orthonormal, has the following components:
\footnote
{We use the Latin indices to identify the different forms in
 the coframe and the Greek indices for coordinate components
 of a differential form. The ordinary summation convection is used.} 
\be
g_{\mu\nu}=\eta_{ab}\vt^a_\mu\vt^b_\nu,
\label{g}
\ee
where $\vt^a_\mu$ are the coordinate components of the differential
 form $\vt^a$ in a local coordinate system $\{x^\mu\}$:
 $\vt^a=\vt^a_\mu dx^\mu$.\\
In the framework of the metric structure the only natural
 covariant objects is the Riemann tensor and its traces -
 the Ricci tensor and the scalar curvature. \\
 However, for the frame structure we need some other natural covariant
 substances, which can be actually constructed using the natural operators
 on the algebra bundle on $M$.
 The most mathematical useful covariant second order differential
 operator is the Hodge - de Rham Laplacian
\footnote{We use the notation of d'Alembertian $\square$ in order
 to emphasize the Lorentzian signature of the manifold.}
\be
\square \ \vt^a=(d*d*+*d*d) \vt^a
\ee
This operator commutes with the exterior derivative operator $d$
 and with the Hodge dual map $*$. In the special case of a flat manifold
 it is the usual Laplace operator (for Euclidean signature)
 and the wave operator - d'Alembertian ( for Lorentzian signature).\\
 The field equation is declared \cite {K-I} in the following form
\be
\square \ \vt^a=\lambda \vt^a,
\label{fe}
\ee
where $\la$ is some function of the frame $\vt^a$ and its exterior
 derivatives.\\ 
The equation (\ref{fe}) represents a system of nonlinear PDE since 
the Laplacian operator $\square$ itself depends on the particular 
choice of the coframe field $\{\vt^a\}$ . \\
Another useful forms of the field equation (\ref{fe}) was proposed
 in \cite{Hehl}:
\begin{equation}\left[  \square + \frac{1}{4} \ensuremath{{}^\star} 
    \left( \vt^\beta \wedge \square \ \eta_\beta \right) \right]
  \eta_\alpha = 0,
  \label{eq:decomp-second}
\end{equation}or, alternatively:
\begin{equation}\left[  \square - \frac{1}{4} 
    \left(e_\beta\hook\square \ \vt^\beta \right) \right]
  \eta_\alpha = 0,
  \label{eq:decomp-second'}
\end{equation}
where $\eta_a=\eta_{ab}\vt^b$ is the down indexed basic 1-form 
and $e_a$ is the  vector field dual to the frame field $\vt^a$.\\
In the special case of a spherical-symmetric static coframe the
 field equation (\ref{fe}) has a unique asymptotic-flat solution.\\
Namely, it is shown \cite{K-I} that the coframe:
\begin{equation}
\vt^0=e^{-\frac{m}{r}}dx^0,\qquad
\vt^i=e^{\frac{m}{r}}dx^i \qquad  i=1,2,3.
 \label{vt1}
\end{equation}
provides a solution of the field equation (\ref{fe}).\\
The correspondent metric is the Yilmaz-Rosen metric:
\begin{equation}
ds^2=e^{-2\frac{m}{r}}dt^2 -e^{2\frac{m}{r}}(dx^2+dy^2+dz^3).
\label{yr}
\end{equation}
This metric is known to be in a good accordance with the 
observation data.\\
The problem of derivation of the field equation (\ref{fe}) 
from some suitable variational principle is discussed in \cite{K-I}
 and\cite{Hehl}.\\
In the present work we generalize the coframe field (\ref{vt1})
 to the following form
\be
\vt^0=e^{-f}dx^0, \qquad
\vt^i=e^fdx^i \qquad  i=1,2,3,
\label{vt2}
\end{equation}
where $f$ is an arbitrary function of coordinates.\\
 It is  reasonable to call this coframe \it a pseudo-conformal coframe\rm. \\
The correspondent metric element will be
\be
ds^2=e^{-2f}dt^2 -e^{2f}(dx^2+dy^2+dz^3).
\label{metric}
\ee
The metric of such a form is known in the classical theory of GR
 as the Majumdar-Papapetrou metrics.
We are looking now for the  conditions on the function $f$,
 under which the coframe field
 (\ref{vt2}) satisfies the field equation (\ref{fe}).
 It turns out that the function $f$ must be spatial
 (elliptically) harmonic.
\section{Pseudo conformal coframe}
\thm
The coframe
\be
\vt^0=e^{-f}dx^0 \qquad \vt^i=e^{f}dx^i \qquad i=1,2,3,
\label{frame}
\ee
where $f=f(t,x,y,z)$ is an arbitrary scalar function on the manifold $M$,
provides the solution of the field equation 
\be
\square \ \vt^a=\lambda \vt^a,
\ee
if and only if one of the following condition are satisfied:
\begin{itemize}
\item The function $f=f(t,x,y,z)$ is static
 (is not depend on the time coordinate $t$) and spatial harmonic 
\begin{equation}
\triangle f=f_{xx}+f_{yy}+f_{zz}=0.
\label{f-har}
\end{equation}
\item The function $f=f(t,x,y,z)$ is homogeneous 
(is not depend on the spatial coordinates $x,y,z$) and linear
\be
f=a t.
\label{f-t}
\ee
\end{itemize}
\ethm
\prf

The straightforward calculations of the Hodge-de Rham Laplacian
 for the coframe (\ref{frame}) give the following expressions 
 (see Appendix 1):
\brn
\square \ \vt^0&=&e^{-2f}(f_{xx}+f_{yy}+f_{zz}+f_x^2+f_y^2+f_z^2)\vt^0
+3e^{2f}(f_t^2+f_{tt})\vt^0\\ 
&&+4(f_{xt}+f_xf_t) \vt^1 +4(f_{yt}+f_yf_t) \vt^2+4(f_{zt}+f_zf_t)\vt^3.
\ern

\brn
\square \ \vt^1&=&\Big[e^{2f}(f_{tt}+3f_t^2)-e^{-2f}(f_{xx}+f_{yy}+f_{zz}- f_x^2-f_y^2-f_z^2)\Big]\vt^1\\
&&+4f_tf_x\vt^0.
\ern

\brn
\square \ \vt^2&=&\Big[e^{2f}(f_{tt}+3f_t^2)-e^{-2f}(f_{xx}+f_{yy}+f_{zz}- f_x^2-f_y^2-f_z^2)\Big]\vt^2\\
&&+4f_tf_y\vt^0.
\ern
\brn
\square \ \vt^3&=&\Big[e^{2f}(f_{tt}+3f_t^2)-e^{-2f}(f_{xx}+f_{yy}+f_{zz}- f_x^2-f_y^2-f_z^2)\Big]\vt^3\\
&&+4f_tf_z\vt^0.
\ern
In accordance with the field equation (\ref{fe}) the non-diagonal 
terms of the Laplacians have to vanish and we obtain two different 
 possibilities
 $$f_t=0$$
 or 
$$f_x=f_y=f_z=0$$.
Let us first consider the static condition: $f_t=0$.\\
The Hodge-de Rham Laplacians remain in the diagonal form
\be
\square \ \vt^0=e^{-2f}(f_{xx}+f_{yy}+f_{zz}+f_x^2+f_y^2+f_z^2)\vt^0,
\ee
\be
\square \ \vt^1=e^{-2f}(-f_{xx}-f_{yy}-f_{zz}+ f_x^2+f_y^2+f_z^2)\vt^1 \ \ \
 etc.
\ee
So the field equation (\ref{fe}) are satisfied if and only if 
\be
\triangle f=f_{xx}+f_{yy}+f_{zz}=0.
\ee
Let us consider now the homogeneous conditions $f_x=f_y=f_z=0$.\\
The function $f$ depends now only on the time coordinate $t$.\\
The Hodge-de Rham Laplacians are
\be
\square \ \vt^0=3e^{2f}(f_{tt}+f_t^2)\vt^0,
\ee
\be
\square \ \vt^1=e^{2f}(f_{tt}+3f_t^2)\vt^1 \ \ \ etc.
\ee
The field equation (\ref{fe}) is satisfied if and only if 
\be
f_{tt}=0
\ee
and the unique solution is $f=a t+b$.\\
The arbitrary constant $b$ can be omitted by the suitable 
re-calibration of the time coordinate and we obtain the
 one-parametric solution $f=at$.
\eprf \\
The coframe (\ref{frame}) corresponds to the metric element
\be
ds^2=e^{-2f}dt^2-e^{2f}(dx^2+dy^2+dz^2).
\ee
The curvature scalar for such a metric with a  static
 harmonic function $f$ is (see calculations in Appendix 2)
\be
R=-8e^{-4f}(f_x^2+f_y^2+f_z^2).
\ee
Note, that scalar curvature scalar is non positive for every 
particular choice of the harmonic function $f$.\\
In the homogeneous case we obtain
\be
R=16f_t^2e^{8f}=16a^2e^{at}
\ee
and the scalar curvature is positive for all finite values of $t$.
\section{Static solutions}
Static solutions of the pseudo-conformal type of the field 
equation (\ref{fe}) are determined by a particular choice 
of a harmonic scalar function $f(x,y,z)$. Note, that the 
scalar potential in the Newton theory of gravity (in vacuum) must
  be a harmonic function as well. So in framework of our model we
 obtain a direct relation between the classical (non-relativistic)
 gravity and the relativistic modification.\\
\it Every physical sensible classical solution of the Newton gravity
 has it's respective counterpart as a solution of the relativistic
 field equation
 (\ref{fe}). \rm
\subsection{Spherical-symmetric solution}
The Laplace equation (\ref{f-har}) in the spherically-symmetry case
 has an unique asymptotically vanishing solution
\be
f=\frac{m}{r},
\ee
where $m$ is an arbitrary scalar constant.
Consequently, we have the pseudo-conformal coframe (\ref{frame})
 - a solution of the field equation (\ref{fe}):
\begin{equation}
\vt^0=e^{-\frac{m}{r}}dx^0 \qquad 
\vt^i=e^{\frac{m}{r}}dx^i, 
\end{equation}
which corresponds to the Yilmaz-Rosen metric:
\begin{equation}
ds^2=e^{-2\frac{m}{r}}dt^2 -e^{2\frac{m}{r}}(dx^2+dy^2+dz^3).
\label{Y-R}
\ee
This solution represents the gravity field of
 a single pointwise body with a mass $m$.\\
The Tailor expansion of the line element (\ref{Y-R})
 up to and including the order $\frac{1} {r^2}$ takes the form
\begin{equation}
ds^2=\Big(1-{\frac{2m}{ r}} + {\frac{2m^2} {r^2}}+...\Big) dt^2 -
 \Big( 1+{\frac{2m}{r}} + {\frac{2m^2} {r^2}}+...\Big) (dx^2+dy^2+dz^2).
\end{equation}
By comparison, the Schwarzchild line element,
 in the isotropic coordinates, is
\begin{equation}
ds^2=\Big[\frac{1-\frac{m}{2r}} {1+\frac{2m} {2r}}\Big]^2dt^2-
\Big(1+\frac{m}{2r}\Big)^4(dx^2+dy^2+dz^2)
\end{equation}
and it's Tailor expansion up to the same order is
\begin{equation}
ds^2=\Big(1-\frac{2m}{ r} + {\frac{2m^2} {r^2}}+...\Big)dt^2 - \Big(1+\frac{2m}{ r}
 +\frac{3}{ 2} {\frac{m^2} {r^2}}+...\Big)(dx^2+dy^2+dz^2).
\end{equation}
The difference between these two line elements is only in the second
 order term of the spatial part and in the third order term of the
 temporal part of the metric.
It is impossible to distinguish, by the latest experiment techniques,
 between these two different line elements.\\  
The curvature scalar for this metric is
\be
R=-2\frac{m^2}{r^4}e^{-2\frac{m}{r}}.
\ee
Note, that the curvature scalar is nonsingular for
 all permissible values of the radius $r$ and vanishes  
only at the origin.

\subsection{Solution with $n$-singular points}
The Laplace equation (\ref{f-har}) is linear so
 any linear combination of solutions provides
 a new solution. Therefore, we have a solution 
with the function $f$ of the following form
\be
f=\sum_{i=0}^n \frac{m_i}{r_i},
\ee
where $m_i$ are arbitrary scalar constants.\\
The coframe field (\ref{frame}) with such a  choice
 of the harmonic function $f$ corresponds to the
 following metric
\begin{equation}
ds^2=e^{-2\sum_{i=0}^n \frac{m_i}{r_i}}dt^2 -
e^{2\sum_{i=0}^n \frac{m_i}{r_i}}(dx^2+dy^2+dzy^3).
\ee
The solution can be interpreted as a metric of a static
 system of $n$ pointwise bodies.\\
Thus, the field equation (\ref{fe}) has a solution with
 a static configuration of masses. 
Note, that the same type of solutions appear in classical
 field theories and in the Einstein gravity (Weyl solution).
\subsection{Solid body solution}
The Newton potential for a classical gravity field 
produced by a solid body (for example - massive ball 
of non-vanishing radius) can be described by an integral
 on the compact 3-dimensional domain:
\be 
f(x,y,z)=\int_V \frac {\rho}{|r-r^\prime|}dV^\prime,
\label{int-har}
\ee
where $\rho=\rho(x,y,z)$ is a local mass density.   \\
The function (\ref{int-har}) is a scalar harmonic function.
So we obtain  a soltion of the field equation (\ref{fe}) 
and consequently a metric element of the prescribed type (\ref{metric}).
For a ball of radius $R$ with a homogeneus distribution of mass we obtain
\be
f=\frac{M}{r},
\ee
where $M$ is the total mass of the ball
$$M=\int_{V^\prime} \rho dV^\prime$$.
Note two remarkable classical results that remain true also in our sheme:
\begin{itemize}
\item The external gravity field of a massive spherical body is equal 
to the field of the point with the mass equal to the mass of the ball 
and located in its center.  
\item The gravity field within a spherical cavity is zero.
\end{itemize}
\subsection{Axial symmetric solution}
The axial symmetric static solution of the Einstein 
equation in vacuum is given by the Weyl metric \cite {W}.
 The metric element can be written as follows \cite {Is}:
\be
ds^2=e^\sigma dt^2-e^{-\sigma}\Big[e^\chi(d\rho^2+dz^2)+\rho^2d\vt^2\Big],
\label{weyl1}
\ee
where $\sigma=\sigma(\rho,z)$ is a harmonic function, 
that is, satisfies the $2D$-Laplace equation
\be
\triangle \sigma=\sigma_{\rho\rho}+\sigma_{zz}+\rho^{-1}\sigma_\rho=0
\label{weyl2}
\ee
and the function $\chi=\chi(\rho,z)$ is given by the following two equations
\be
\chi_z=\rho\sigma_\rho\sigma_z,
\label{weyl3}
\ee
\be
\chi_\rho=\frac{1}{2}\rho(\sigma_\rho^2-\sigma_z^2).
\label{weyl4}
\ee
The consistency of the last two equation is guaranteed by (\ref{weyl2}).\\
Note, that the equations (\ref{weyl3}-\ref{weyl4}) can be solved for an
 every particular choice of a harmonic function $\sigma$.\\
Let us return to the field equation (\ref{fe}).
The axial symmetric static solution can be given by the pseudo conformal frame:
\be
\vt^0=e^{-\frac{\sigma}{2}}dt \qquad \vt^i=e^{\frac{\sigma}{2}}dx^i,
\ee
where $\sigma$ is a harmonic function and in two dimensional case 
satisfies the equation (\ref{weyl2}).\\
The resulting metric is 
\be
ds^2=e^{-\sigma} dt^2-e^{\sigma}\Big(d\rho^2+dz^2+\rho^2d\phi^2\Big).
\label{ax-m}
\ee
Note that the metric (\ref{ax-m})  has the same form as the metric 
of Weyl with a vanishing function $\chi$. So instead of the system 
(\ref{weyl2},\ref{weyl3},\ref{weyl4}) we have only one equation (\ref{weyl2}).
Let us consider two particular solutions of the equation (\ref{weyl2}).\\
It is easy to see that the function
\be
\sigma=\frac{m_1}{\sqrt{\rho^2+(z-a)^2}}+\frac{m_2}{\sqrt{\rho^2+(z+a)^2}}
\ee
satisfies this equation.\\
This function has two singular points on $z$-axis at $z=-a$ and $z=a$, 
thus it represents the gravity field of two massive pointwise bodies
 located at these points.\\
Another particular class of solutions of the equation (\ref{weyl2})
 can be given by the following harmonic function $\sigma$ \cite {Is}
\be
\sigma=\delta \ln\Big(\frac{z-a+R^{(-)}}{{z+a+R^{(+)}}}\Big)=
\delta \ln\Big(\frac{R^{(-)}+R^{(+)}-2a}{{R^{(-)}+R^{(+)}+2a}}\Big),
\label{rod}
\ee
where $\delta$ is a dimensionless constant and $a$ is a constant
 with dimension of length. The functions $R^{(\pm)}$ are
\be
R^{(\pm)}=\Big[\rho^2+(z\pm a)^2\Big]^{\frac12}.
\ee
In the framework of the Newtonian theory the function $\sigma$ of a 
type (\ref{rod}) represents the gravity potential of an infinitesimally 
thin uniform rod with a density proportional to $\delta$ and with
 a length equal to $2a$. The center of the rod is in the origin and its
 ends lying on $z$-axis at $z=-a$ and $z=a$.
For such a choice of the function $\sigma$ the metric element is
\be
ds^2=\Big(\frac{{R^{(-)}+R^{(+)}+2a}}{R^{(-)}+R^{(+)}-2a}\Big)^\delta dt^2 - \Big(\frac{R^{(-)}+R^{(+)}-2a}{{R^{(-)}+R^{(+)}+2a}}\Big)^\delta\Big(d\rho^2+dz^2+\rho^2d\phi^2\Big)
\ee
\section{Homogeneous solution}
The second choice of the function $f=at$ in the theorem gives a homogeneous solution depending on the time coordinate. The corresponding metric element is
\be
ds^2=e^{-2at}dt^2 - e^{2at}(dx^2+dy^2+dz^2).
\ee
The curvature scalar is (see Appendix)
\be
R=18a^2e^{2at}.
\ee
Observe, that for the negative values of the arbitrary parameter
 $a$ the curvature scalar is equal to $18a^2$ for $t=0$ and
 vanishes for $t \to \infty$.
 So the solution describes a world that is expanding exponential
 from a finite radius of order $\frac{1}{a}$ to infinity with
 the time coordinate $t$. \\
For $a>0$ we have an evolution of an inverse type and this solution
 can not be consistent with the observations.\\
Using a new time coordinate 
\be
\tau=\pm \frac{1}{a}e^{-at}
\ee
the metric element can be rewritten as
\be
ds^2=d\tau^2 - \frac{1}{a^2\tau^2}(dx^2+dy^2+dz^2).
\ee
The time $\tau$ is the proper time at each point in the space.

\section{The theory of Yilmaz}
Yilmaz \cite {yilmaz58}, \cite {yilmaz76} was presented a theory of 
gravitation in which the basic dynamical variable is a scalar field $\phi$.
 The metric tensor $g_{\mu\nu}$ is not an independent dynamical variables,
 but a function of $\phi$.
The field equations of the theory are the following ones:\\
the Einstein field equation
\be
R^\mu_\nu-\frac12 \delta^\mu_\nu R=8\pi T^\mu_\nu
\label{fey1}
\ee
and the Laplace equation for the scalar field
\be
g^{\mu\nu}\phi_{;\mu\nu}=0
\label{fey2}
\ee
The tensor $T^\mu_\nu$ is the usual energy-momentum tensor 
for the scalar field
\be
T^\mu_\nu=\frac{1}{8\pi}(2g^{\mu\lambda}\phi_{,\nu}\phi_{,\lambda}-
\delta^\mu_\nu g^{\lambda\tau}\phi_{,\lambda}\phi_{,\tau})
\ee
The main result of the Yilmaz approach is as follows:\\
The field equation (\ref{fey2}) can be solved by the following 
special form of the metric tensor
\be
g_{00}=e^{-2\phi};\qquad \qquad g_{ii}=-e^{2\phi},\qquad i=1,2,3
\ee
where $\phi$ is scalar function of spatial coordinates $x,y,z$.\\
In this case the field equation  (\ref{fey1}) satisfies automatically
 and the equation (\ref{fey2}) reduces to the Newtonian equation
\be
\triangle \phi= \Big(\frac{\partial^2}{\partial x^2}
+\frac{\partial^2}{\partial y^2}
+\frac{\partial^2}{\partial z^2}\Big) \phi=0
\ee
The unique asymptotically vaniching harmonic function with 
one singular point produce the metric of Yilmaz, 
which is in a good accordance with the observations.\\
The analysis above shows that the magic result of Yilmaz reproduces 
in the pseudo-conformal sector of solutions of the field equation (\ref{fe}).

\vskip 0.3true cm
{\bf Acknowledgments.} The author is very grateful to Professor
 F.W. Hehl for his excellent hospitality and stimulation discussions
 in the University of Cologne.
I would like also to thank Professor S. Kaniel for constant support
 and many useful discussions.
\appendix
\section{Calculations of Laplacians}
Let us calculate the Hodge-de Rham Laplacian for the coframe (\ref{frame}).\\
The exterior derivative of the differential form $\vt^0$ is
\begin{eqnarray*}
d\vt^0&=&e^{-f}dt\wedge (f_x dx+f_y dy+f_z dz)\\
&=&e^{-f}\vt^0\wedge (f_x \vt^1+f_y\vt^2+f_z \vt^3).\\
\end{eqnarray*}
The Hodge dual of this expression is 
 \begin{eqnarray*}
*d\vt^0&=&e^{-f}(f_x\vt^2\wedge\vt^3-f_y\vt^1\wedge\vt^3
+f_z \vt^1\wedge \vt^2)\\
&=&e^{f}(f_x dy\wedge dz-f_y dx\wedge dz+f_z dx\wedge dy).\\
\end{eqnarray*}
The exterior derivative of this expression is 
\footnote {We use the following notation $\vt^{123}=\vt^{1}\wedge\vt^{2}\wedge\vt^{3}$ e.t.c.}
  \begin{eqnarray*}
d*d\vt^0&=&e^{f}\Big[(f_{xx}+f_{yy}+f_{zz}+f_x^2+f_y^2+f_z^2)
dx\wedge dy\wedge dz\\
&+&(f_{xt}+f_xf_t) dt\wedge dy\wedge dz -(f_{yt}+f_yf_t) dt\wedge dx\wedge dz\\
&+&(f_{zt}+f_zf_t)dt\wedge dx\wedge dy\Big]\\
&=&e^{-2f}(f_{xx}+f_{yy}+f_{zz}+f_x^2+f_y^2+f_z^2)\vt^{123}\\
&+&(f_{xt}+f_xf_t) \vt^{023} -(f_{yt}+f_yf_t) \vt^{013}+(f_{zt}+f_zf_t)
\vt^{012}.
\end{eqnarray*}
Taking the Hodge dual we have
   \begin{eqnarray*}
*d*d\vt^0&=&e^{-2f}(f_{xx}+f_{yy}+f_{zz}+f_x^2+f_y^2+f_z^2)\vt^0\\
&+&(f_{xt}+f_xf_t) \vt^1 +(f_{yt}+f_yf_t) \vt^2+(f_{zt}+f_zf_t)\vt^3.
\end{eqnarray*}
The Hodge dual of the differential form $\vt^0$ is
$$
*\vt^0=\vt^{123}=e^{3f}dx\wedge dy\wedge dz.
$$
The exterior derivative of this expression is
  $$
d*\vt^0=3e^{3f}f_t dt\wedge dx\wedge dy\wedge dz=3e^{f}f_t\vt^{0123}.
$$
The Hodge dual is
 $$
*d*\vt^0=3e^{f}f_t.
$$
The exterior differential is
\brn
d*d*\vt^0&=&3e^{f}\Big[(f_t^2+f_{tt})dt+(f_tf_x+f_{tx})dx+(f_tf_y+f_{ty})dy\\
&+&(f_tf_z+f_{tz})dz\\
&=&3e^{2f}(f_t^2+f_{tt})\vt^0+3(f_tf_x+f_{tx})\vt^1+3(f_tf_y+f_{ty})\vt^2\\
&+&3(f_tf_z+f_{tz})\vt^3.
\ern
The Hodge-de Rham Laplacian of the differential form $\vt^0$ is
\brn
\square \ \vt^0&=&e^{-2f}(f_{xx}+f_{yy}+f_{zz}+f_x^2+f_y^2+f_z^2)\vt^0
+3e^{2f}(f_t^2+f_{tt})\vt^0\\ 
&+&4(f_{xt}+f_xf_t) \vt^1 +4(f_{yt}+f_yf_t) \vt^2+4(f_{zt}+f_zf_t)\vt^3.
\ern
Let us calculate the Hodge-de Rham Laplacian of the differential form $\vt^1$.
\brn
d\vt^1&=&d(e^fdx)=e^f(f_tdt\wedge dx-f_ydx\wedge dy-f_zdx\wedge dz)\\
&=& e^ff_t\vt^{01}-e^{-f}f_y\vt^{12}-e^{-f}f_z\vt^{13}.
\ern
The Hodge dual of this expression is
\brn
*d\vt^1&=&e^ff_t\vt^{23}+e^{-f}f_y\vt^{03}-e^{-f}f_z\vt^{02}\\
&=&e^{3f}f_t dy\wedge dz+e^{-f}f_y dt\wedge dz-e^{-f}f_z dt\wedge dy.
\ern
The exterior derivative of this expression is
\brn
d*d\vt^1&=&e^{3f}(f_{tt}+3f_t^2)dt\wedge  dy\wedge dz+e^{3f}(f_{tx}+3f_tf_x)dx \wedge dy\wedge dz\\
&&+e^{-f}(f_{xy}-f_xf_y)dx\wedge dt\wedge dz +e^{-f}(f_{yy}-f_y^2)dy\wedge dt\wedge dz \\
&&-e^{-f}(f_{xz}-f_xf_z)dx \wedge dt\wedge dy-e^{-f}(f_{zz}-f_z^2)dz \wedge dt\wedge dy\\
&=&e^{2f}(f_{tt}+3f_t^2)\vt^{023}+(f_{tx}+3f_tf_x)\vt^{123}-e^{-2f}(f_{xy}-f_xf_y)\vt^{013}\\
&&e^{-2f}(f_y^2-f_{yy})\vt^{023}+e^{-2f}(f_xf_z-f_{xz})\vt^{012}-e^{-2f}(f_z^2-f_{zz}) \vt^{023}.
\ern
Consequently
\brn
*d*d\vt^1&=&\Big[e^{2f}(f_{tt}+3f_t^2)-e^{-2f}(f_{yy}-f_y^2)-e^{-f}(f_{zz}-f_z^2)\Big]\vt^1\\
&&+(f_{tx}+3f_tf_x)\vt^0 +e^{-2f}(f_{xy}-f_xf_y)\vt^2+e^{-2f}(f_{xz}-f_xf_z)\vt^3.
\ern
For the dual form we have
$$
*\vt^1=\vt^{023}=e^fdt\wedge dy\wedge dz.
$$
The exterior derivative of this expression is
\brn
d*\vt^1&=&-e^{f}f_xdt \wedge dx \wedge dy\wedge dz=-e^{-f}f_x\vt^{0123}.
\ern
The Hodge dual is
\brn
 *d*\vt^1&=&-e^{-f}f_x.
\ern
The exterior derivative is
\brn
d*d*\vt^1&=&-e^{-f}(f_{xx}-f_x^2)dx-e^{-f}(f_{xy}-f_xf_y)dy-e^{-f}(f_{xyz}-f_xf_z)dz\\
&&-e^{-f}(f_{xt}-f_xf_t)dt\\
&=&-(f_{xt}-f_xf_t)\vt^0-e^{-2f}(f_{xx}-f_x^2)\vt^1-e^{-2f}(f_{xy}-f_xf_y)\vt^2\\
&&-e^{-2f}(f_{xyz}-f_xf_z)\vt^3.
\ern
The Hodge-de Rham Laplacian of the differential form $\vt^1$ is
\brn
\square \ \vt^1&=&\Big[e^{2f}(f_{tt}+3f_t^2)-e^{-2f}(f_{xx}+f_{yy}+f_{zz}- f_x^2-f_y^2-f_z^2)\Big]\vt^1\\
&&+4f_tf_x\vt^0.
\ern
For the Hodge-de Rham Laplacian of the differential forms $\vt^2$ and $\vt^3$ the expressions are similar:
\brn
\square \ \vt^2&=&\Big[e^{2f}(f_{tt}+3f_t^2)-e^{-2f}(f_{xx}+f_{yy}+f_{zz}- f_x^2-f_y^2-f_z^2)\Big]\vt^2\\
&&+4f_tf_y\vt^0.
\ern
\brn
\square \ \vt^3&=&\Big[e^{2f}(f_{tt}+3f_t^2)-e^{-2f}(f_{xx}+f_{yy}+f_{zz}- f_x^2-f_y^2-f_z^2)\Big]\vt^3\\
&&+4f_tf_z\vt^0.
\ern
\section{Curvature}
For calculation of the curvature tensor we use the formulas  \cite {L_L} for non-vanishing components of Ricci tensor in the case 
of a ``diagonal'' metrics.\\
Let the components of the metric tensor will be as following 
\footnote{In all formulas above the summation over repeated 
indices is not used.}
\be
g_{ii}=e_ie^{2F_i}, \qquad g_{ij}=0 \qquad  for \qquad i\ne j
\ee
where $e_i=\pm 1$.\\
The components of Ricci tensor are:
\be
R_{ik}=\sum_{l\ne i,k}\Big(F_{l,k}F_{k,i}+F_{i,k}F_{l,i}-F_{l,i}F_{l,k}-F_{l,i,k}\Big) \qquad {for} \qquad i\ne k
\ee
\br
R_{ii}&=&\sum_{l\ne i}\Big [F_{i,i}F_{l,i}-F_{l,i}^2-F_{l,i,i} \nonumber\\
&+&e_ie_le^{2(F_i-F_l)}\Big(F_{l,l}F_{i,l}-F_{i,l}^2-F_{i,l,l}-F_{i,l}\sum_{m\ne i,l}F_{m,l}\Big)\Big]
\er
Thus we have for the pseudo conformal metric (\ref{metric})

\be
R_{00}=-6f_t^2-3f_{tt}-e^{-4f}\triangle f
\ee
\be
R_{11}=e^{4f}(4f_t^2-f_{tt})-2f_x^2-\triangle f
\ee
\be
R_{22}=e^{4f}(4f_t^2-f_{tt})-2f_y^2-\triangle f
\ee
\be
R_{33}=e^{4f}(4f_t^2-f_{tt})-2f_z^2-\triangle f
\ee
The curvature scalar is
\be
R=18f_t^2e^{2f}-2e^{-2f}(\triangle f+f_x^2+f_y^2+f_z^2)
\ee
Static solution satisfying the equation $\triangle f=0 $ gives
\be
R=-2e^{-2f}(f_x^2+f_y^2+f_z^2).
\ee
Note, that the curvature scalar is non-positive.\\
For the homogeneous solution we have
\be
R=18f_t^2e^{2f}
\ee
Note, that the curvature scalar is nonnegative.

\end{document}